\input lecproc.sty
\input epsf.tex
\contribution{On Black Hole Spins and Dichotomy of Quasars}
\contributionrunning{On Black Hole Spins and Dichotomy of Quasars}
\author{Rafa{\l} Moderski@1@2, Marek Sikora@2 and Jean-Pierre Lasota@1}
\authorrunning{Moderski et al.}
\address{@1UPR 175 du CNRS; DARC, Observatoire de Paris, Section de
Meudon, 92195~Meudon, FRANCE
@2Nicolaus Copernicus Astronomical Center, Bartycka~18,
00-716~Warszawa, POLAND}
\abstract{
Most quasars are known to be radio quiet and according to the ``spin
paradigm'', which connects their radio-loudness with the value of the
black hole spin, they must harbor very slowly rotating black holes. On
the other hand, quasars are powered by accretion which tends to
increase the spin of the black hole. We show that the Blandford-Znajek
mechanism is not efficient enough to counteract the spinning up of
black holes by an accretion disk, regardless of the accretion rate. We
establish conditions under which it is possible to obtain low values
of the black hole spin, in the scenario involving a sequence of many
accretion events with random angular momenta. Our results are used to
select possible evolutionary scenarios which can explain radio
bimodality of quasars in terms of the ``spin paradigm''.}
\titlea{1}{Introduction}
One of the most intriguing properties of quasars is the large range
and bimodal distribution of their radio-loudness.  Quasars can
approximately be divided into two classes: radio-loud (RL) quasars and
radio-quiet (RQ) quasars, with the RL-quasars having the ratio of
their radio to optical fluxes $F_{\rm 5GHz}/F_{\rm B} \ge 10$
(Kellerman 1989). RL objects are rather rare (they consist only about
$10 \%$ of the AGN population) and they almost always reside in
elliptical galaxies.  Despite the differences in radio-loudness, the
optical and ultraviolet spectra of RL and RQ objects look very similar
(Francis et al.~1993; Zheng et al.~1997). Since the optical-UV
radiation is produced by accretion flows, the above indicates that
accretion conditions do not differ significantly in these two classes
of quasars. This supports the idea that the parameter which determines
the power of jets and radio-loudness of quasars is the black hole spin
(Blandford~1990). This, so called `spin paradigm', is supported by
recent calculations of parameters of Galactic
`micro-quasars'. Considering the thermal emission and frequency of
quasi-periodic oscillations in X-ray binaries, Zhang et al.~(1997)
have found that galactic sources with strong jet activities harbour
fast-rotating black holes.

The spin paradigm seems to be in odds with the interpretation of broad
red wings of fluorescent iron line in RQ Seyfert galaxies in terms of
reflection model in the Kerr metric (Iwasawa et al.~1996). However, as
was recently shown by Reynolds \& Begelman~(1997), the ``Kerr''
profiles of the Fe K$\alpha$ line can be mimicked by radiation
scattered by the matter contained in the region between the marginally
stable orbit and the horizon of a non-rotating black hole.

The main purpose of this paper is to verify the basic assumption of
the ``spin paradigm'', which is that the population of quasars on the
whole is dominated by objects with low spin black holes. We
investigate whether such an assumption can be reconciled with any
reasonable evolutionary scenario of super-massive black holes.

\titlea{2}{Equilibrium Spin}
We assume simple Kerr geometry in which a black hole in expressed in
the terms of its energy-mass, M, and angular momentum, J.  Evolution
of the black hole is described by the set of equations (Moderski \&
Sikora~1996a):
$$ c^2 {{\rm d}M \over {\rm d}t}= e_{\rm in} \dot {\cal M} - P, \eqno(1)$$
$$ {{\rm d}J \over {\rm d}t}=j_{\rm in} \dot {\cal M} -
{ P \over \Omega_{\rm F} }, \eqno(2)$$
$$ P \simeq {1\over 8} {B^2 r_h^4 \over c} \Omega_F (\Omega_h - \Omega_F),
\eqno(3) $$
where $\dot {\cal M} \equiv {\rm d} {\cal M} / {\rm d}t$ is the
accretion rate, $e_{\rm in}$ and $j_{\rm in}$ are specific energy and
angular momentum of matter at inner edge of the accretion disc, $P$ is
the power extracted by the Blandford--Znajek (B-Z) mechanism,
$\Omega_{\rm F}$ is the angular velocity of the magnetic field lines
threading the horizon, $\Omega_h$ is the angular velocity of a black
hole, and $B$ is the intensity of the magnetic field on the horizon.
 
Using dimensionless units: $A=cJ/GM^2 \equiv J/J_{max}$, $\tilde
j_{in} = cj_{in}/GM$, $\tilde e_{in} = e_{in}/c^2$ and $\tilde \Omega
= GM\Omega/c^3$, and noting that $\tilde \Omega_h = A/(2\tilde r_h)$,
we rewrite Eqs. (1), (2) and (3) as
$$ {dA \over dt} = {1 \over Mc^2}
\left (\dot {\cal M}c^2 (\tilde j_{in} - 2 A \tilde e_{in}) 
- P \left ({2 \tilde r_h \over kA} -2A \right )\right ) , \eqno (4) $$
$$ {d \ln {M} \over dt} = {1\over Mc^2} (\dot {\cal M} \tilde e_{in} - P),
\eqno (5) $$
$$ P \simeq {k(k-1)\over 32} {G^2\over c^3} A^2 \tilde r_h^2 B^2 M^2 \, ,
\eqno (6) $$
where $k=\Omega_F/\Omega_h$.  Assuming that the pressure of the black
hole magnetic field, $B^2/8\pi$ is balanced by the ram pressure of the
innermost parts of an accretion flow,
$$ p \sim \rho c^2 \sim {c \dot {\cal M} \over 4 \pi r_h^2}    , \eqno(7) $$
we obtain 
$$P = {\pi \over 4} k(k-1) {G m_p c\over \sigma_T} \dot m A^2 M
= {k(k-1)\over 16}A^2 \dot m L_{Edd}, \eqno(8) $$
where $\dot m = {\cal M}c^2/L_{Edd}$, and $L_{Edd}= 4 \pi G m_p c
M/\sigma_T$.

From Eqs. (4) and (8) one can calculate the equilibrium spin $A_{eq}$ for
which $dA/dt=0$. This spin does not depend on accretion rate, and for the
maximum efficiency of the B-Z mechanism, i.e. for $k=1/2$,
$A_{eq}=0.997$. Because this value is so high evolution of a black
hole for any $A<0.9$ is very well approximated by equation
$$ {dA \over d\ln M} = {(\tilde j_{in} - 2 A \tilde e_{in}) \over
\tilde e_{in}}
\, \eqno(9) $$
which is obtained by dividing Eq. (4) over Eq. (5) and assuming $P=0$.
Eq. (9) has analytical solution for $r_{in} =r_{ms}$ (Bardeen~1970)
and for $r_{in} = r_{mb}$ (Abramowicz \& Lasota~1980, Moderski \&
Sikora~1996b), where $r_{ms}$ is the marginally stable orbit and
$r_{mb}$ is the marginally bound orbit. In particular, it can be shown
that initially nonrotating black hole with mass $M_0$, after accreting
$\Delta m$ from the disk, will be spun-up to $A \sim \Delta m /
M_0$. This shows that only those super-massive black holes which
accrete from the disk much less than their initial mass can avoid
spinning up to high values of $A$.
\titlea{3}{Switching Between Pro- and Retrograde Accretion Discs}
Matter, accreting from the disc which rotates in the opposite
direction than the black hole, carries negative angular momentum and
reduces the spin of the black hole. We call such a process a
`retrograde accretion'. This process was shown by Moderski \&
Sikora~(1996b) to be very a efficient mechanism of spinning down a
black hole. It suffices to accrete $\sim 0.2$ of the initial black
hole mass, $M_0$, to decelerate the black hole from its maximum spin,
$A_{max}=1$, to zero.

Of course, an accretion of larger amount of matter than $0.2 M_0$ will
spin-up a black hole again, and in order to maintain the low value of
the time-averaged spin, the evolution of a black hole must be governed
by many ``small'' ($\delta m \ll M_0$) accretion events. This case is
illustrated in Fig. 1, where we show the evolution of the black hole
spin for $\delta m = 0.01 M_0$. The evolution illustrated on this
Figure was obtained neglecting the B-Z mechanism, with an assumption
that accreting matter forms geometrically thin disc, and that the
angular momentum of the disc is randomly switched between two opposite
directions.
\begdoublefig 0 cm
{\centerline{\epsfysize=6.3truecm\epsfbox{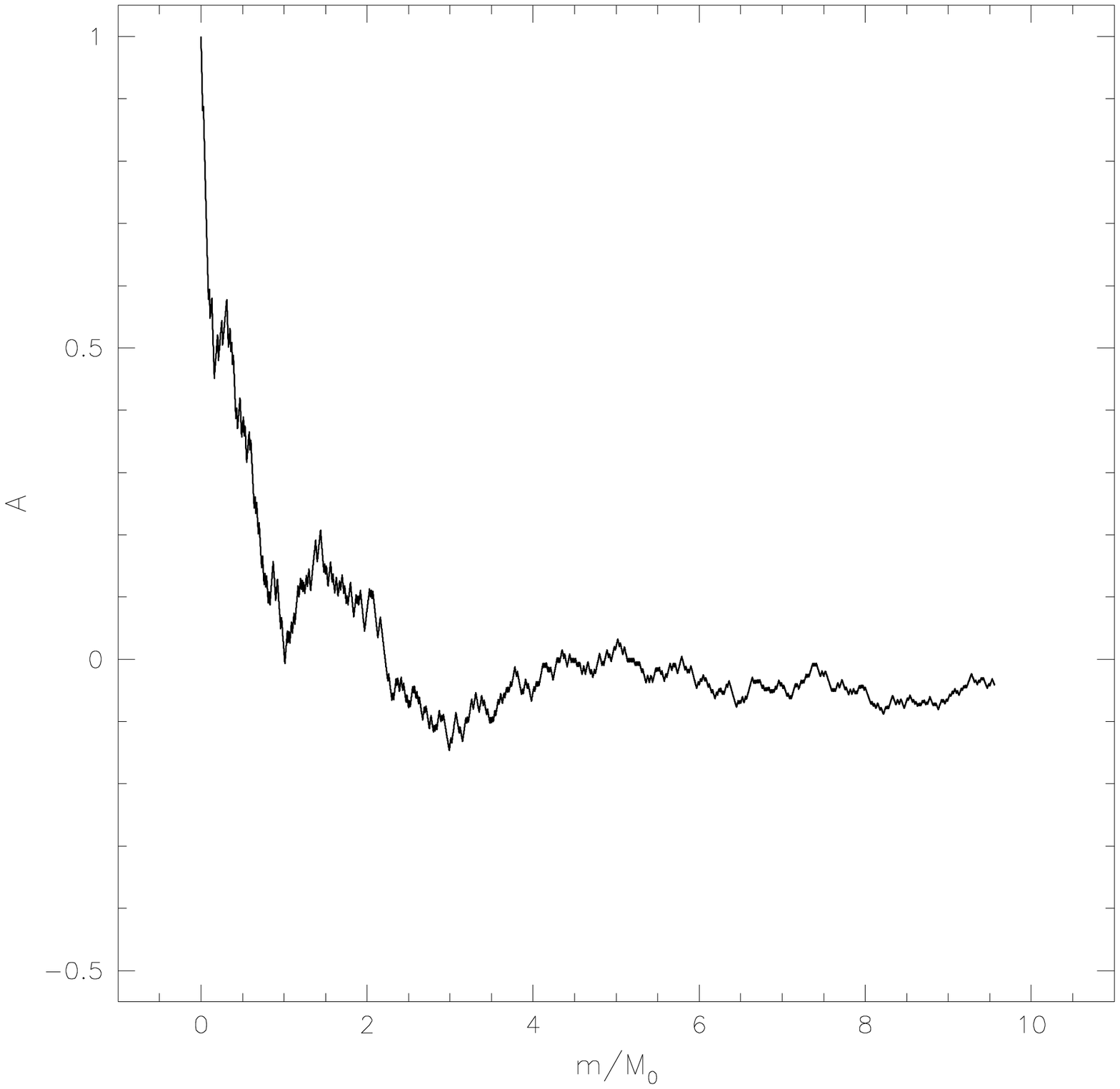}}
\figure{1}{
Example of the evolution of the black hole's spin. Black hole evolves
from the maximally rotating state and accretes each time a portion
$0.01$ of its initial mass.
}}
{\centerline{\epsfysize=6.3truecm\epsfbox{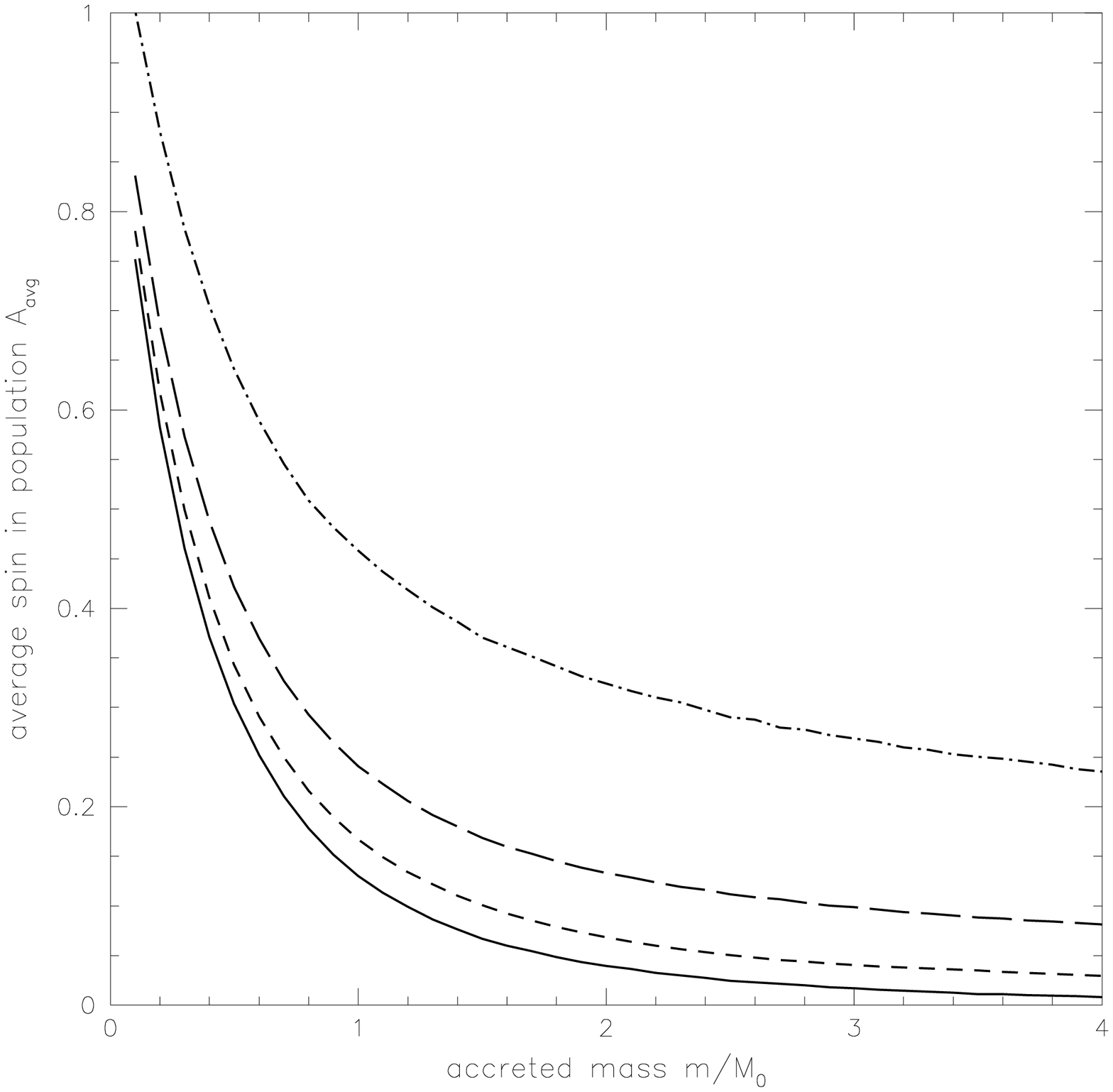}}
\figure{2}{
Average spin and spread in the population of black hole evolving from
$A_0=1$ as a function of accreted mass. Average spin is mark\-ed with
solid line while $1\sigma$ standard deviation in distribution
(assuming Gaussian shape) is mark\-ed as dotted, dashed and long dashed
lines for $\delta m=0.001, 0.01$ and $0.1 M_0$ respectively. Only
$+1\sigma$ contours are plotted
}}
\enddoublefig
Fig.~1 shows that after the phase of deceleration, the spin of a black
hole fluctuates around zero.  We also checked behavior of the
population of black holes evolving from the $A_0=1$ by performing a
number of `numerical evolutions' for various $\delta m$. We have found
that the final average spin in the population does not depend on
$\delta m$ but only on the total accreted mass. The average spin of
the population and standard deviation of the distribution are plotted
versus total accreted mass in Fig. 2. Note that results presented in
Fig. 1 and 2 hardly depend on the B-Z mechanism, because, as shown in
\S2, this mechanism is dynamically efficient only for $A\sim 1$.
\titlea{4}{Discussion}

The fact that the equilibrium spin of a black hole is very close to
the maximal value and does not depend on the accretion rate, implies
that evolution of black holes with $A<0.9$ is strongly dominated by
accretion.  Taking this into account, and assuming that radio-loudness
of quasars is related to the black hole spin, we discuss two possible
accretion history scenarios which can eventually lead to the observed
radio dichotomy of quasars.  The first one (model `A') is based on the
assumption that during the entire history of a given quasar, the
direction of the rotation of its accretion disk remains constant.  The
observed excess of RQ-quasars over RL-quasars can then be explained in
terms of the spin paradigm only if the quasar population is dominated
by objects containing a low spin black hole. Thus, a majority of black
holes must be initially formed with a very low spin and the amount of
matter accreted from the disk must be much smaller than the initial
mass of the black hole. The second scenario (model `B') is based on
the assumption that the history of a quasar can be described by a
sequence of accretion events with randomly oriented angular momentum
vectors.  In this case black holes can be formed with high spin,
provided that after their formation, they accrete enough amount of
matter to be slowed down to a low spin value.

To quantify the above constraints, we estimate a value of the black
hole spin, $A_c$, which corresponds to the radio-to-optical flux ratio
$F_{5{\rm GHz}}/F_B \sim 10$ dividing quasars into the RL and RQ
objects.  Assuming that the fraction of jet power converted to
radiation is $10$\%, and that bolometric correction for jet radiation
at $\nu_R \sim 5$ GHz is of the same order as that for the accretion
disk radiation in the B-band, and noting that $\nu_R/\nu_B \sim
10^{-5}$, we find that $F_R/F_B \simeq 10$ corresponds to $P/L_d
\simeq 10^{-3}$, where $L_d$ is the bolometric luminosity of the
accretion disk.  Dividing $P$ given by Eq. (8) by $L_d =(1-\tilde
e_{in})\dot {\cal M}c^2$, we find
$${P\over L_d} \simeq {k(1-k)\over 16}{A^2 \over (1 - \tilde e_{in})} ,
\eqno (10) $$
and, therefore,
$$ A_c \simeq  \sqrt {16 \times 10^{-3} (1 -\tilde e_{in}) \over k(1-k)} \, .
\eqno (11)$$
For the maximum efficiency of the Blandford-Znajek mechanism,
i.e. $k=1/2$, and for radiation efficiency of a disk $(1 -\tilde
e_{in}) \sim 0.1$, Eq. (11) gives $A_c \simeq 0.1$.

For $A_c \simeq 0.1$, the model `A' is viable provided that $90$\% of
quasars are born with $A < 0.1$ and that the mass of the matter
accreted from the disk is less than $0.1$ of the initial mass of the
black hole.  The RL-quasars can therefore be the objects where:

\leftline {(a) - a black hole is already formed with the high ($A > A_c$) 
spin, and/or \hfill}
\leftline {(b) - a black hole accreted from the disk more than $10$\% of its 
initial mass, and/or \hfill} 
\leftline {(c) - a black hole coalesces with another  black hole having 
a comparable mass. \hfill} 

Depending on whether the supermassive black holes are formed mostly
from the gas cloud or from stellar cluster they may acquire higher or
lower initial spin. In the giant ellipticals, where formation of the
nucleus is much faster and presumably involves larger amounts of gas
than in spirals, the growth of a black hole can be dominated by a
collapse of a gas cloud. A formation of a black hole from a gas cloud
is most likely accompanied by a formation of a massive disk. High
accretion rate from such a disk can well support strong magnetic
fields in rapidly rotating black holes and, together with a jet
powered by the rotating black hole, it can create the phenomenon of
the RL-quasars.  This high accretion phase is limited by the amount of
matter left in the disk after the black hole formation process, and
later, when the accretion rate drops, an advection dominated disk can
be formed (Rees et al.~1982, Narayan \& Yi~1995, Abramowicz et
al.~1995). In such a disk, the radiation efficiency is very low, and
since the spin of the black hole does not change, the radioludness,
$P/L_d$, reaches very high values. These objects are good candidates
for FRI radiogalaxies.

If all black holes are formed with low spin, the option (b) or (c)
applies. In this case about $10$\% of black holes must be spun-up in
later evolutionary phases. High values of a spin can be reached by a
coalescence of supermassive black holes, as proposed by Wilson \&
Colbert (1995), or by accretion of more than $100$\% of their initial
mass from the disk. These two scenarios are expected to follow a
merger of two galaxies, at least one of them gas rich.  The total mass
accreted during $10^8$ years (what is a typical life-time of
RL-quasars deduced from the radio spectra) with the rate $10 \,
M_{\odot}$ yr$^{-1}$ (required to produce observed UV luminosities) is
enough to spin-up black holes with masses $\le 10^9 M_{\odot}$ up to
$A \sim 1$.

For the model `B', black holes can be spun-down and maintained with a
low value of a spin by an accretion disk which changes direction of
the rotation.  Thus, the population of quasars can be dominated by
objects containing black holes with a spin $A < A_c$, even if their
initial spin is high.  However, as one can deduce from Fig. 2, the
total accreted mass, required to spin-down a black hole to $A<A_c
\simeq 0.1$, is $> 1M_0$, and number of accretion events required to
maintain a low spin is larger than $100$.  Such accretion events may
be due to matter delivered to the very central region by molecular
clouds.  Spinning down the black hole can also be provided by
capturing black holes with masses $10^6 M_{\odot}$, which are
predicted to be formed in recombination era due to Jeans instability
of the primary condensations (Loeb 1993).  In the model `B', the
RL-quasar phenomenon can be related to coalescence of black holes
and/or to giant accretion events induced by galaxy mergers (as in the
model `A'), as well as to the formation of black holes with high
initial spin.

The above considarations have implications on the range of
radio-loudness of RL-quasars.  For $0.1 \le A < 1$ and for the maximum
efficiency of the B-Z mechanism, the ratio $P/L_d$ spans the range
$10^{-3} - 10^{-1}$, which corresponds to the range of the
radio-to-optical flux ratio $10-10^3$ covered by most RL-quasars
(Bischof and Becker 1997).  There is, however, a number of RL-quasars
with the radio-to-optical flux ratio corresponding to $0.1 < P/L_d
\simeq 1$, and such high radio-loudnesses require super-Eddington
accretion rates, with $\dot m$ in the range $10-100$ (note that for
super-Eddington accretion rates $L_d \simeq const = L_{Edd}$ and,
therefore $P/L_d \propto \dot m$). Somewhat smaller $\dot m$ is
allowed, if one takes into account that for quasars with the highest
ratio $L_R/L_B$, the intrinsic value of $L_B$ is higher than observed
due to the extinction in the quasar (Baker~1997).

Finally, we would like to emphasize that basic constraints on
accretion history of quasars, imposed by the assumption that the
radioloudness is scaled by the square of the black hole spin, depend
on how strong is the magnetic field of the black hole that can be
supported by an accretion flow. Unfortunately, this problem has not
yet been solved.  In this paper, we assumed that the maximum magnetic
field is limited by the ram pressure of accretion flow very closely to
the horizon.  This leads to a higher value of $A_{eq}$ than in the
case when the magnetic pressure is balanced by radiation or gas
pressure in the disk (Moderski \& Sikora~1996a; 1997). However, for
radiation pressure dominated disks, the equilibrium spin $A_{eq} >0.1$
and our main conclusions remain the same.

\smallskip

RM and MS acknowledge the support of the KBN grant 2P03D01209.

\begrefchapter{References}
\ref Abramowicz, M.A., Chen, X., Kato, S., Lasota, J.-P. \& Regev, O.,
1995, ApJ, 438, L37
\ref Abramowicz, M.A. \& Lasota, J.-P., 1980, Acta Astronomica, 20, 35
\ref Baker, J.C., 1997, MNRAS, 286, 23
\ref Bardeen, J.M., 1970, Nature, 226, 64
\ref Bischof, O.B. \& Becker, R.H., 1997, AJ, 113, 2000
\ref Blandford, R.D., 1990, in Active Galactic Nuclei, ed. T. J.-L. Courvoisier
\& M. Mayor (Saas-Fee Advanced Course 20) (Berling-Springer), 161
\ref Francis, P.J., Hooper, E.J.,\& Impey, C.D., 1993, AJ, 106, 417
\ref Iwasawa, K., et al., 1996, MNRAS, 282, 1038
\ref Kellerman, K.I., Sramek, R., Schmidt, M., Shaffer, D.B., \&
Green, R., AJ, 1989, 1195
\ref Loeb, A., 1993, ApJ, 403, 542
\ref Moderski, R. \& Sikora, M., 1996a, MNRAS, 283, 854
\ref Moderski, R. \& Sikora M., 1996b, A\&AS, 120C, 591
\ref Moderski R. \& Sikora, M., 1997, INTEGRAL Workshop Proceedings
\ref Narayan R. \& Yi I., 1995, ApJ, 452, 710
\ref Rees, M.J., Begelman, M.C., Blandford, R.D. \& Phinney, E.S.,
1982, Nature, 295, 17
\ref Reynolds, C.S. \& Begelman, M.C., 1997, ApJ, in press
\ref Wilson, A.S. \& Colbert, E.J.M., 1995, ApJ, 438, 62
\ref Zhang, S.N., Cui, W., \& Chen, W., 1997, ApJ, 482, L155
\ref Zheng, W., Kriss, G.A., Telfer, R.C., Grimes, J.P., \& Davidsen, A.F.,
1997, ApJ, 475, 496
\endref

\byebye